\documentclass{article}

\begin{document}

\title{Scalar nature of the nuclear density functional}

\author{B. G. Giraud \\
bertrand.giraud@cea.fr, Institut de Physique Th\'eorique, \\
DSM, CE Saclay, F-91191 Gif/Yvette, France}

\date{\today} 
\maketitle

\begin{abstract}

Because of the rotational invariance of the nuclear Hamiltonian, there exists 
a density functional (DF) for nuclei that depends only on two scalar 
densities. Practical calculations boil down to radial, one-dimensional ones.

\end{abstract}

\bigskip
Let $Z,N,A \equiv Z+N$ be the proton, neutron and mass numbers, respectively. 
The nuclear Hamiltonian $H$ is known to be invariant under rotations. 
Therefore, besides $Z$ and $N,$ nuclear ground states (GSs) carry  good 
quantum numbers, $J$ and $M,$ for the total angular momentum and its 
$z$-component. Two cases occur : either $J=0,$ hence the GS is not degenerate
and its density is isotropic, or $J>0,$ hence one sees a trivial degeneracy 
for a magnetic multiplet of GSs, the densities of which, non isotropic, 
contain multipole components \cite{Doba} \cite{GirPLB} up to order $2J,$  
with the same monopole for all members of the multiplet. 

In any case, the ensemble density operator for the GS(s) of a given nucleus,
\begin{equation}
{\cal B}_{ZN}=(2J+1)^{-1} \sum_{M} |ZNJM \rangle \langle ZNJM |,
\label{ensmbl}
\end{equation}
is a scalar under rotations. In any case also, the GS energy obtains by,
\begin{equation}
E_{ZN}={\rm Tr}\ {\cal B}_{ZN}\, H.
\end{equation}

Every spherical harmonic function, except a monopole, integrates out to zero.
Therefore, only the monopole components of the proton and neutron densities
contribute to the density normalizations which identify a nucleus,
\begin{equation}
\int_0^{\infty} r^2 dr\, \rho_{p0}(r)=Z,\ \ \ \ 
\int_0^{\infty} r^2 dr\, \rho_{n0}(r)=N,
\label{idntnucl}
\end{equation}
the position $r$ being here taken as a scalar rather than a vector. 
In the following, for conciseness, we omit the subscript $0$ and often denote 
$\rho$ the pair $\{\rho_p,\rho_n\}.$

Consider the ``density constrained search'' \cite{Levy} \cite{Lieb} 
for a minimal energy,
\begin{equation}
{\rm Inf}_{\rho \rightarrow \{Z,N\}} \left[ \left(\, {\rm Inf}_{{\cal B} 
\rightarrow \rho}\, {\rm Tr}\ {\cal B}\, H\, \right) + \int_0^{\infty} r^2 
dr\,  [\, u_p(r)\, \rho_p(r) + u_n(r)\, \rho_n(r)\, ] \right],
\end{equation}
where it is understood that the many-body density operator ${\cal B}$ is 
restricted to be a scalar under rotations. Also $u_p$ and $u_n$ are scalar. 
With two distinct densities $\rho_p, \rho_n,$ two potentials 
are needed to represent the external potential used by the Hohenberg-Kohn 
\cite{HK} theorem as a functional Lagrange multiplier to constrain the 
density. Notice that now we do not put subscripts $Z,N$ to ${\cal B},$ 
because the normalizations, Eqs. (\ref{idntnucl}), are implemented at the 
stage of the ``outer'' minimization. At the ``inner'' stage, $N$ and $Z$ do 
not need to be integers.

This inner minimization, ${\rm Inf}_{{\cal B} \rightarrow \rho},$ defines
a density functional,
\begin{equation}
F[\rho] \equiv {\rm Inf}_{{\cal B} \rightarrow \rho}\ {\rm Tr}\ {\cal B}\, H,
\label{nucfuncl}
\end{equation}
and, in that sector defined by the additional constraints, 
Eqs. (\ref{idntnucl}), with now $Z$ and $N$ physical integers, the GS energy 
of a nucleus results from,
\begin{equation}
E_{ZN}={\rm Inf}_{\rho}\, F[\rho],\ \ \ \ 
\int_0^{\infty} r^2 dr\, \rho_p(r)=Z,\ \ \ \ 
\int_0^{\infty} r^2 dr\, \rho_n(r)=N.
\label{GSnrg}
\end{equation}
It seems, therefore, that Eqs. (\ref{nucfuncl}) and (\ref{GSnrg}) provide the 
basis of a fulfledged nuclear DF theory (DFT) in radial space, a symmetrized 
theory {\it \`a la} G\"orling \cite{Goer1}.

However, one must first remove an ambiguity in the definition of the degree of 
freedom $r$ and the associated density $\rho(r).$ For $A$ nucleons, the 
simplest set of degrees of freedom are the single nucleon coordinates 
$\vec r_1,$ $\vec r_2,$ ... $\vec r_A$ and the simplest definition of the 
density consists in integrating out all of them but one,
\begin{equation}
\rho(\vec r)=A \int d\vec r_1\, d\vec r_2\, ...\, d\vec r_{A-1}\  
|\Psi(\vec r_1,\vec r_2,...,\vec r_{A-1},r)|^2\, .
\label{lab}
\end{equation}
But, as discussed by \cite{EngBar}, it is more physical to use a density 
$\sigma \left(\vec r-\vec R \right),$ measured from the 
center-of-mass (CM) coordinate $\vec R=(\vec r_1+...+\vec r_A)/A$ of the 
nucleus, rather than the density $\rho(\vec r),$ defined in the laboratory 
frame. Since $H$ is also translation invariant, the wave function for $A$ 
nucleons is rather an ``internal'' one, 
$\psi_{int}(\vec \xi_1,...,\vec \xi_{A-1}),$ of $(A-1)$ Jacobi coordinates 
only,
$ \vec \xi_1=\vec r_2-\vec r_1,\, $
$ \vec \xi_2=\vec r_3-(\vec r_2+\vec r_1)/2,\, $ ... ,
$ \vec \xi_{A-1}=\vec r_A-(\vec r_{A-1}+\vec r_{A-2}+...+\vec r_1)/(A-1)\, .$

It turns out that the last Jacobi coordinate is proportional to 
$\vec r_A - \vec R,$ namely, 
$\vec \xi_{A-1}=\frac{A}{A-1}\, (\vec r_A-\vec R).$
Except for trivial scaling factors, the ``internal'' density appears naturally 
to be,
\begin{equation}
\sigma(\vec \xi)=A \int d\vec \xi_1\, d\vec \xi_2\, ...\, d\vec \xi_{A-2}\  
|\psi_{int}(\vec \xi_1,\vec \xi_2,...,\vec \xi_{A-2},\vec \xi)|^2\, 
.
\label{int}
\end{equation}
Strictly speaking, what we called $F[\rho]$ should rather be a functional
of $\sigma.$ Unfortunately, there is no need to stress that calculations with 
Jacobi coordinates are much more complicated than calculations in the 
laboratory frame.

The solution found in \cite{Gircm} to link internal density $\sigma$ and 
laboratory density $\rho$ consists in trapping the CM by a harmonic potential. 
The Hamiltonian becomes,
\begin{equation}
{\cal H}=\sum_{i=1}^A \frac{p_i^2} {2m} + 
\frac{K m}{2}\, \left(\sum_{i=1}^A \vec r_i \right)^2 + 
\sum_{i>j=1}^A v_{ij}\, + \sum_{i>j>k=1}^A w_{ijk}.
\label{hamil}
\end{equation}
Here $v$ is the usual two-body interaction, and one can also include a 
three-body interaction $w$ or even the luxury of more-body ones. Except for 
the trap, Galilean invariance is requested; no density dependence in $v,w,...$ 
is allowed. 

The trap is parametrized by an arbitrary, but fixed constant $K,$ to be 
chosen for maximum convenience of practical calculations. The term,
$K m \left(\sum_{i=1}^A \vec r_i \right)^2,$ can also be written as, 
$A m\, \omega^2 R^2,$ with the CM coordinate 
$\vec R=A^{-1} \sum_{i=1}^A \vec r_i $ and $\omega=\sqrt{K A};$ 
the CM frequency depends on the mass number $A.$ The same form also 
shows that the trap is the sum of a one-body and a two-body 
operators, representable in second quantization without coefficients 
depending on $A,$
\begin{equation}
\frac{K\, m}{2}\, \left(\sum_{i=1}^A \vec r_i \right)^2 = \frac{K\, m}{2}\, 
\left(\, \sum_{i=1}^A r_i^2 + 2\, \sum_{i>j=1}^A \vec r_i \cdot \vec r_j \, 
\right). 
\end{equation}

In each sector specified by integer $Z$ and $N,$ the GS(s) of 
${\cal H}$ factorize(s) as product(s) of a common Gaussian $\Gamma$ for the 
CM and internal wave function(s) of the $(A-1)$ Jacobi coordinates,
\begin{equation}
\Psi(\vec r_1,...,\vec r_A)=\Gamma(R)\ 
\psi_{int}(\vec \xi_1,...,\vec \xi_{A-1}),\ \ 
\Gamma(R)=
\pi^{-\frac{3}{4}}\, b^{-\frac{3}{2}}\, \exp\left[-\frac{R^2}{2b^2}\right],
\end{equation}
with $b=\left[\, \hbar/(A\, m\, \omega)\, \right]^{\frac{1}{2}}.$
The Gaussian is rotation invariant. Its does not perturb the physical 
quantum numbers $J,M$ when ${\cal H}$ is substituted for $H.$ 

As shown in \cite{Gircm} the link between $\rho$ and $\sigma$ is a trivial, 
invertible convolution,
\begin{equation}
\rho(\vec r) = \frac{A^3}{(A-1)^3} \int d\vec R\ [\Gamma(R)]^2\ 
\sigma\left[\frac{A}{A-1}\, \left(\vec r-\vec R\right)\right].
\label{convol} 
\end{equation}
This link is the same for any member of a magnetic 
multiplet. Hence it extends to the scalar densities $\rho$ and $\sigma$ 
provided by that kind of ensemble described by Eq. (\ref{ensmbl}). The vector 
coordinates we used temporarily to handle the non isotropic densities of 
individual members of a magnetic multiplet can be reduced again to scalars. 
The convolution, $\rho = \Gamma^2 * \sigma,$ see Eq. (\ref{convol}), actually 
becomes,
\begin{equation}
r\, \rho(r) = \frac{2\, {\pi}^{-\frac{1}{2}} A^3}{(A-1)^3\, b} \int_0^{\infty} 
ds\, \exp\left[-\frac{r^2+s^2}{b^2}\right]\, 
\sinh\left[\frac{2rs}{b^2}\right]\, s\ \sigma\left[\frac{A}{A-1}\, s \right].
\label{scalconvol} 
\end{equation}

Since it is easy to second quantize ${\cal H}$ and use the laboratory proton 
and neutron densities, 
$\rho_p(r)={\rm Tr}\ {\cal B}\ c^{\dagger}_{pr}\, c_{pr}\, ,$\
$\rho_n(r)={\rm Tr}\ {\cal B}\ c^{\dagger}_{nr}\, c_{nr}\, ,$
where, with obvious notations, we have introduced proton and neutron creation 
and annihilation operators at a scalar position $r,$ traces upon scalar 
density operators can be calculated in Fock space. Then one only tunes 
Eqs. (\ref{GSnrg}) into,
\begin{equation}
E_{ZN}+\frac{1}{2} \hbar \sqrt{K A}={\rm Inf}_{\rho}\, {\cal F}[\rho],\ \  
\int_0^{\infty} r^2 dr\, \rho_p(r)=Z\, ,\ \ 
\int_0^{\infty} r^2 dr\, \rho_n(r)=N\, .
\label{GSnrgcm}
\end{equation}

We conclude by claiming that a DFT for nuclei energies (and every other scalar)
is available with {\it scalar} densities. Practical designs of a nuclear DF, 
${\cal F}[\rho] \equiv 
{\rm Inf}_{{\cal B} \rightarrow \rho} {\rm Tr}\, {\cal B}\, {\cal H},$ 
simultaneously valid for doubly even, odd and doubly odd nuclei, can be 
attempted in the laboratory radial frame.

From a DFT, one expects densities and energies. At the cost of ignoring
density multipoles\footnote{The wave function multipoles, nonetheless,
contribute to the density monopole.} 
other than monopoles, our approach does provide formally {\it exact} energies,
in a most simplified, one-dimensional theory. A major problem of nuclear 
physics is the prediction of exotic nuclei. Our result is of a special 
interest for the study of the neutron drip line, where neutron halos are 
notoriously difficult to describe. Dimensional reduction should make it easier
to better focus the theory on the design\footnote{See \cite{Gir} for 
expansions of densities in polynomials constrained by matter conservation.} 
of the functional ${\cal F},$ a still formidable problem. Whether the same 
reduction to ``radial pictures'' might also simplify the non local \cite{Gil}
versions of the DFT, in particular the quasi-local versions derived from 
Skyrme force models and labelled ``energy density'' theories, is likely.

\medskip \noindent
{\it Acknowledgement}: A discussion with J. Dobaczewski is gratefully 
acknowledged.

\end{document}